\documentclass[10pt,aps.prl,twocolumn,superscriptaddress,floatfix]{revtex4-1} 

\usepackage{amsmath}
\usepackage{graphics}
\usepackage{color}
\usepackage{caption}
\usepackage{subcaption}
\usepackage{float}
\usepackage{array}
\usepackage[margin=1.0in]{geometry}
\usepackage{ulem}

\begin{document}
\title{The Effect of Experimental Parameters on Optimal Transmission of Light Through Opaque Media}
\author{Benjamin Anderson$^1$, Ray Gunawidjaja$^1$, and Hergen Eilers$^{*}$}
\address{Applied Sciences Laboratory, Institute of Shock Physics, Washington State University,
Spokane, WA 99210-1695}
\date{\today}

\email{eilers@wsu.edu}

\begin{abstract}
Spatial light modulator (SLM) controlled transmission of light through opaque media is a relatively new experimental method with wide applications in various fields. While there has been a surge in research into the technique there has been little work reported considering the effects of various experimental parameters on the efficiency of optimization.  In this study we explore the effects of various experimental conditions on optimization and find that the intensity enhancement depends on the number of modulated channels, number of phase steps, feedback integration radius, beam spot size, and active SLM area. We also develop a model, based on the propagation of a Gaussian beam with a random phase front, to account for most of the measured effects.

\vspace{1em}

%42.25.Dd	Wave propagation in random media,  42.25.Bs	Wave propagation, transmission and absorption,  42.25.Fx	Diffraction and scattering, 42.25.-p	Wave optics,   42.15.Eq	Optical system design,  42.70.Jk	Polymers and organics

\end{abstract}

%\vspace{3em}

%\bibliographystyle{osajnl}
%\bibliography{ASLbib}

\maketitle

\vspace{1em}

\section{Introduction}
Scattering materials (e.g. paper, paint, clouds, etc.) are generally viewed as hindrances to optical propagation.  Light propagating through such a materials behaves in a diffusive manner with scattering resulting in the amplitude and phase patterns being destroyed \cite{Sebbah01.01}.  However, in 1990 Freund theoretically showed that if you correctly shape the wavefront of the beam incident on the scattering system, the opaque system can be used as a lens or other high precision optical device \cite{Freund90.01}.  

The first experimental realization of an opaque lens was demonstrated by Vellekoop and Mosk \cite{Vellekoop07.01}.  Their technique involved using a liquid crystal on silicon (LCOS) spatial light modulator (SLM) to change the phase of the incident wavefront such that the modulated beam matches the transmission eigenmodes of the scattering sample. This leads to the light being focused through the sample onto a target area  \cite{Vellekoop07.01,Vellekoop08.01} .   This method of wavefront control -- using an LCOS SLM -- has promising applications for astronomical and biological imaging \cite{Mosk12.01,Stockbridge12.01}, flourescene microscopy \cite{Vellekoop10.02,Wang12.01}, sub-diffraction limit focusing \cite{Vellekoop10.01, Putten11.01},  focusing and compression of ultrashort pulses \cite{Katz11.01,McCabe11.01}, spectral filtering \cite{Small12.01,Park12.01,Paudel13.01,Beijnum11.01}, and light polarization control \cite{Park12.02,Guan12.01}.

Our interest in the technique of SLM controlled optimal transmission is as a mechanism for verifying physically unclonable functions (PUFs) \cite{Goorden13.01}.  PUFs are materials with a large number of random degrees of freedom that are practically impossible to recreate due to their inherent randomness \cite{Goorden13.01}.  An example of a PUF is a scattering system, such as a nanoparticle (NP) doped polymer.  In such a PUF the positions of all the NPs are the degrees of freedom and since there are trillions of randomly distributed particles in the scattering volume it is impossible to recreate the PUF; such that the scattering signature is identical.  While the scattering signature (speckle pattern) is one way of authenticating/characterizing a scattering PUF, another possibility is to use SLM controlled transmission.  For this application the SLM can be used to create a phase profile (challenge) that produces a specific transmission profile (key).  If someone tampers with the PUF the coupling between 
the phase profile and transmission profile will be broken, giving evidence of tampering. 

In order to use a SLM controlled transmission system for secure authentication of scattering PUFs we must first understand how different experimental parameters affect the system.  We therefore build a SLM controlled transmission setup and characterize -- both theoretically and experimentally --  the system's optimization dependence on five different system parameters: SLM bin size, $b$, number of SLM phase steps, $M$, active SLM area, $L^2$, detector integration radius, $r$, and the on-sample beam spot size, $w$.

\begin{figure*}
\centering
 \begin{subfigure}[b]{0.45\textwidth}
 \centering
\scalebox{0.75}{\includegraphics{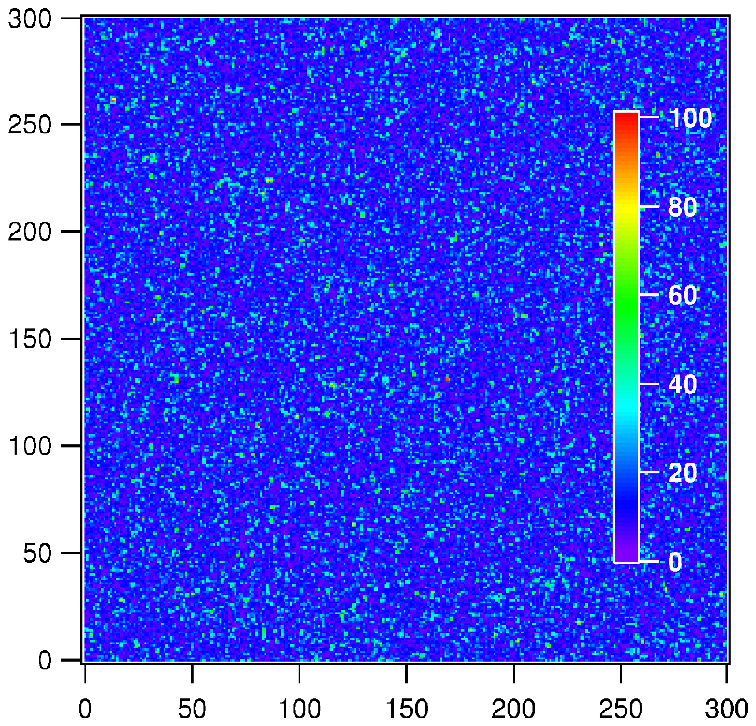}}
\caption[a]{Before}
\end{subfigure}
\qquad
 \begin{subfigure}[b]{0.45\textwidth}
 \centering
 \scalebox{0.75}{\includegraphics{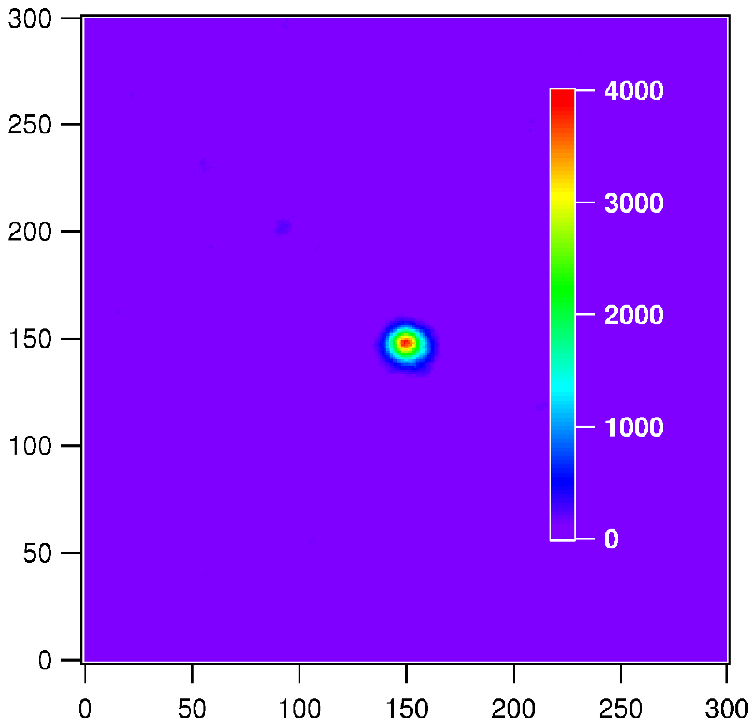}}
 \caption[b]{After}
\end{subfigure}
\caption{(Color Online) Camera image before (a) and after (b) optimization. The image before optimization is dim and random, while afterwards the beam is focused into a tight spot. \textcolor{black}{The camera pixel size is 5.2 $\mu$m and the exposure time is 2 ms for the dim image and 0.127 ms for the optimized image.}}
\label{fig:comp}
\end{figure*}

For comparing the influence of the different variables on optimization we calculate/measure the intensity enhancement, 
$\eta$, which is defined as:

\begin{equation}
\eta \equiv \frac{I}{\langle I_0\rangle},
\label{eqn:etadef}
\end{equation}
where $I$ is the average intensity in the target spot after optimization and $\langle I_0\rangle$ is the ensemble averaged intensity in the target before optimization \cite{Vellekoop07.01}.  Figure \ref{fig:comp} shows an example of the intensity profile both  before-optimization and after-optimization.  The pattern before optimization is a random speckle pattern, while afterwards the pattern is a focused spot.  By calculating the average intensity in the spot both before and after optimization we calculate the enhancement using Equation \ref{eqn:etadef}.

The intensity enhancement due to SLM phase modulation has previously been modeled using an analog to electron conduction \cite{Mello88.01,Beenakker97.01,Pendry92.01,Garcia89.01}, which predicts that the enhancement depends on the number of modulated SLM channels, $N$, as \cite{Vellekoop07.01}:

\begin{equation}
\eta=\frac{\pi}{4}(N-1)+1.
\label{eqn:Vmodel}
\end{equation}
From Equation \ref{eqn:Vmodel} we see that the conduction model predicts that the enhancement should depnd only on the number of modulated SLM channels, with other system/sample parameters having no influence \cite{Vellekoop07.01,Vellekoop08.02}. In practice, however, experiments on a variety of systems are found to give drastically different enhancements for similar $N$ values \cite{Vellekoop07.01, Popoff10.01,Cui11.01,Conkey12.01,Park12.02,Conkey12.02,Guan12.01}.  To account for these variations Yilmaz and coworkers developed a model to include detector noise into the optimization scheme \cite{Yilmaz13.01}.  Taking noise into account, Equation \ref{eqn:Vmodel} transforms into
\begin{equation}
\eta=\frac{\pi}{4}N\left(1-\frac{N}{R^2}\right),
\label{eqn:Ymodel}
\end{equation}
where $R$ is the signal-to-noise ratio of the system.  

From Yilmaz and coworkers model, the enhancement should only depend on the number of SLM channels and the signal-to-noise ratio of the system.  However, in our current work we find that the enhancement depends on the five system variables mentioned earlier, as well as the samples themselves.  Given that our experimental results deviate from Equations \ref{eqn:Vmodel} and \ref{eqn:Ymodel} we propose a beam propagation model to account for the measured enhancement's dependence on system parameters.  The model--which we label the random phase Gaussian beam model (RPGBM)-- is based on the propagation of a beam with a spatially random phase distribution and a Gaussian amplitude distribution.  \textcolor{black}{From both the model and experiment we determine phenomenological equations to describe the enhancement's dependence on different system variables.}

\section{Model}

\subsection{Theory}
The random phase Gaussian beam model (RPGBM) treats the scattering sample as a ``black box'', with the effect of scattering to introduce a random phase pattern to a Gaussian beam.  We begin by assuming a TEM$_{00}$ Gaussian beam incident on the sample, with the beam waist being located at the incident surface.  The incident electric field is therefore

\begin{align}
 E_i(x,y)=E_0e^{-(x^2+y^2)/\sigma_0^2}
\end{align}
where $E_0$ is the incident field strength and $\sigma_0$ is the beam's Gaussian width.  To model scattering, we let the beam width increase, $\sigma_0\rightarrow \sigma$, and introduce a random phase profile, $\Phi(x,y)$. With these transformations the field exiting the sample is given by: 

\begin{align}
 E(x,y)=E_0e^{-(x^2+y^2)/\sigma^2-i\Phi(x,y)}.
 \label{eqn:efint}
\end{align}
\textcolor{black}{We note here that this model of scattering is a simplistic approximation and does not reflect light propagation in real scattering media.  However, due to it's simplistic nature it allows us to perform a wide range of calculations, which are otherwise unfeasible.}

Assuming that the distance from sample to detector, $Z$, is much greater than the beam width, $\sigma<<Z$, we can use Fraunhoffer diffraction theory to determine the beamprofile at the detector.  In Fraunhoffer diffraction theory the diffracted electric field, $E_d(x,y)$ \footnote{ Note that we use primed coordinates to denote the detector plane, and unprimed coordinates for the sample plane.}, is the Fourier Transform of the initial electric field:

\begin{align}
 E_d(x',y')=\int_{-\infty}^{\infty}\int_{-\infty}^{\infty}dxdyE_0\exp\bigg\{-\frac{(x^2+y^2)}{\sigma^2} \nonumber
 \\-i\Phi(x,y)+i\frac{k}{Z}(xx'+yy')\bigg\},
\end{align}
where $k$ is the wavenumber given by $k=2\pi/\lambda$, with $\lambda$ being the wavelength of light. From the diffracted electric field we calculate the beam profile given by

\begin{align}
 I_d(x',y')=\left|E_d(x',y')\right|^2.
 \label{eqn:int}
\end{align}

We model the optimization process by introducing a phase shift, $\psi(x,y)$, to the beam in the sample plane, such that $\psi(x,y)$ represents the influence of SLM phase modulation.  By systematically varying $\psi(x,y)$ according to an optimization algorithm we can reproduce the same process used experimentally.  With the addition of the SLM phase shift, the diffracted field becomes:

\begin{align}
 E_d(x',y')=\int_{-\infty}^{\infty}\int_{-\infty}^{\infty}dxdyE_0\exp\bigg\{-\frac{(x^2+y^2)}{\sigma^2} \nonumber
 \\-i\Phi(x,y)+i\frac{k}{Z}(xx'+yy')+i\psi(x,y)\bigg\}.
 \label{eqn:modfield}
\end{align}
Using Equations \ref{eqn:modfield} and \ref{eqn:int} we can therefore model the detector feedback signal as we change $\psi(x,y)$.

While Equation \ref{eqn:modfield} uses continuous Fourier Transforms, we use discrete Fourier transforms when performing computations. Discretizing Equation \ref{eqn:modfield}  gives

\begin{align}
E_{d; n'm'}=\sum\limits_{n=0}^{N-1}\sum\limits_{m=0}^{N-1} E_0\exp\bigg\{-\frac{(n^2+m^2)\Delta x^2}{\sigma^2} \nonumber
\\-i\Phi_{n,m}+i\frac{2\pi}{N}(nn'+mm')+i\psi_{n,m}\bigg\}.
 \label{eqn:ddiff}
 \end{align}
where we have substituted the $x,y$ coordinates with integer values $n,m$ such that:

\begin{align}
x=n\Delta x && x'=n'\Delta x' \nonumber
\\ y=m\Delta x && y'=m'\Delta x' \nonumber
\end{align}
with $\Delta x$ being the grid spacing in the sample plane and $\Delta x'$ is the grid spacing in the target plane given by

\begin{align}
\Delta x'=\frac{2\pi Z}{Nk\Delta x}.
\label{eqn:uncertainty}
\end{align}
Note that Equation \ref{eqn:uncertainty} implies an inverse relationship between distances in the sample and detector planes.

\subsection{Computational Details}

For our calculations we define a $1000 \times 1000$ grid with an isotropic grid spacing of $\Delta x$, such that $\Delta x<<\sigma$, where $\sigma$ is the Gaussian width of the electric field. Using a random number generator each grid point is assigned a phase value between 0 and 2$\pi$ with the generated numbers having a uniform probability distribution.  The combination of the phase value, $\Phi_{n,m}$, and Gaussian width, $\sigma$ defines the electric at the sample plane as

\begin{align}
E_{n,m}=\sqrt{\frac{2\Delta x^2}{\pi \sigma^2}}\exp\bigg\{-\frac{(n^2+m^2)\Delta x^2}{\sigma^2}\nonumber
\\-i\left(\Phi_{n,m}-\psi_{n,m}\right)\bigg\}
\label{eqn:compfield}
\end{align}
where $\psi_{n,m}$ comes from the SLM modulation and the peak field is defined as:
\begin{align}
E_0=\sqrt{\frac{2\Delta x^2}{\pi \sigma^2}},
\end{align}
such that the total integrated intensity is unity.

Using the sample-plane electric field from Equation \ref{eqn:compfield} we model optimization as follows:
\begin{enumerate}
\item $\psi_{n,m}$ is varied according to the chosen optimization algorithm and parameters.  In this study we use a sequential bin-by-bin optimization method in which one bin is modulated at a time to find the optimal phase value, after which that phase is fixed for that bin.
\item Given $\psi_{n,m}$, $E_{d;n',m'}$ is calculated by taking the DFT of $E_{n,m}$
\item The intensity is found using $E_{d;n',m'}$ and Equation \ref{eqn:int} and the average intensity, $\langle I\rangle$, in a target area of radius $r$ is calculated.  
\item A random number, $\sigma_I$, which represents detector noise, is added to the calculated average intensity with the random numbers having an average magnitude of  $\langle |\sigma_I|\rangle =\sqrt{\langle I\rangle}$.
\item The intensity with noise term, $\langle I\rangle+\sigma_I$, is then used as the feedback signal for the optimization algorithm. 
\end{enumerate}
While we use a sequential bin-by-bin optimization algorithim in this study, the model's optimization steps can easily be adapted for more complex algorithms, such as partitioning \cite{Vellekoop08.01} and genetic optimization \cite{Guan12.01,Conkey12.02}.

\subsection{Model Results}
\subsubsection{Bin Size}
\begin{figure}
\centering
\includegraphics{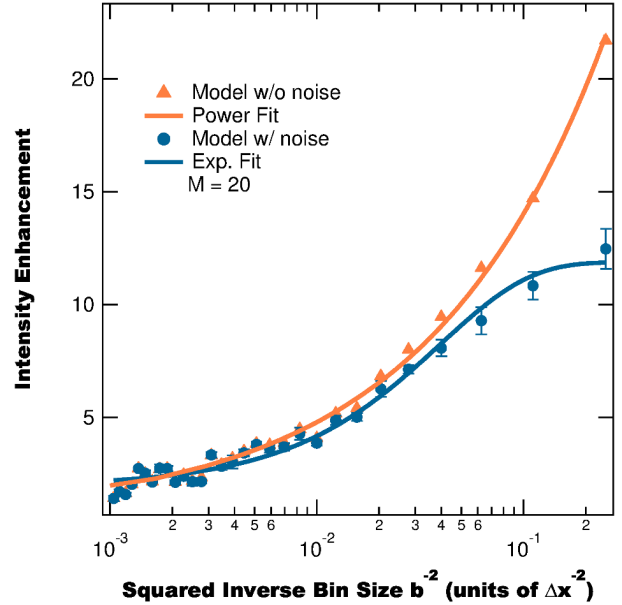}
\caption{(Color Online) Modeled intensity enhancement as a function of inverse squared bin spacing. Without noise the enhancement follows a power function, while with noise the enhancement behaves as an exponential.}
\label{fig:modelbinnoise}
\end{figure}

We first model the effect of bin size, $b$, on optimization by using an integration radius of 2$\Delta x'$ and three different numbers of phase steps, $M=\{2,3,20\}$.  The calculations are performed both with and without noise with the result that the enhancement as a function of bin size depends on whether noise is present or not.  Figure \ref{fig:modelbinnoise} shows a comparison of the enhancement as a function of inverse squared bin size for a calculation with and without noise.  Without noise, the intensity enhancement is found to be proportional to a power function $ (b^{-2})^p$ where $p<1$.  However, when including noise in the calculation we find that the intensity enhancement follows an exponential function,
\begin{align}
\eta=1+\eta_0\exp\left\{-\left(\frac{\alpha b_0}{b}\right)^2\right\},
\label{eqn:modelbin}
\end{align}
where $1+\eta_0$ is the asymptotic enhancement and $\alpha b_0$ is a shape factor, with $\alpha$ being a factor related to the active area of the SLM. Since the results of our calculations depend on whether or not noise is included, the remainder of our calculations we will include the effect of noise; as we can not completely eliminate noise experimentally.

\begin{figure}
\centering
\includegraphics{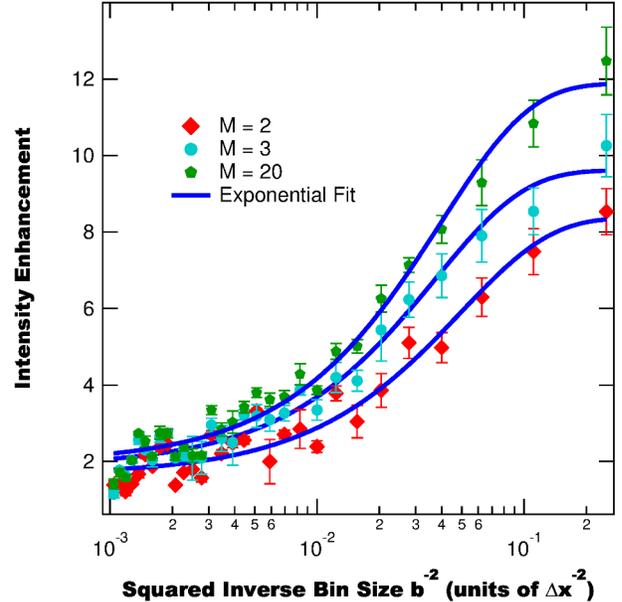}
\caption{(Color Online) Modeled intensity enhancement as a function of inverse squared bin spacing for different $M$ values.  The enhancement follows an exponential function with the amplitude changing with $M$ while the shape parameter remains constant.}
\label{fig:modelbin}
\end{figure}

We next consider how changing the number of phase steps, $M$, influences the parameters in Equation \ref{eqn:modelbin}.  Figure \ref{fig:modelbin} compares the enhancement for different number of phase steps, which we fit to Equation \ref{eqn:modelbin}. From our fits we find that the asymptotic enhancement increases as the number of phase steps increases, while the shape parameter remains constant.

\subsubsection{Active SLM Area}
\begin{figure}
\centering
\includegraphics{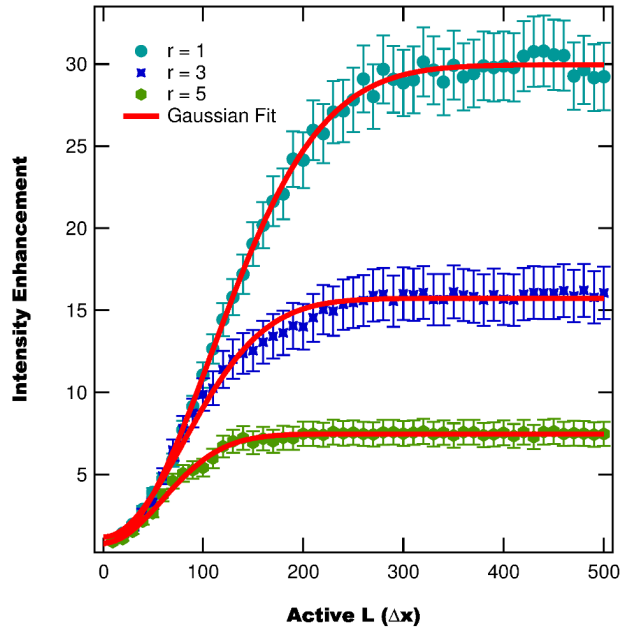}
\caption{(Color Online) Modeled intensity enhancement as a function of active SLM side length.  The enhancement is found to follow a Gaussian function.}
\label{fig:modcrop}
\end{figure}

In the previous section we calculated the effect of changing bin size on optimization.  This represents the first parameter which determines the total number of controllable channels. The other parameter responsible for the total number of bins is the active SLM area, $L^2$, with $L$ being the active SLM side length.  For modeling the effect of changing the active SLM area we use $M=10$ phase steps, a bin size of $b=1$ $\Delta x$, and three different radii: 1 $\Delta x'$, 2 $\Delta x'$, and 5 $\Delta x'$. We calculate the enhancement as a function of active side length, shown in Figure \ref{fig:modcrop}, and find that it behaves as a Gaussian function:

\begin{equation}
\eta=1+\eta_0\left[1-\exp\left\{\left(\frac{L}{\beta \Delta L}\right)^2\right\}\right],\label{eqn:Lmodel}
\end{equation}
where $1+\eta_0$ is the asymptotic enhancement and $\beta\Delta L$ is the Gaussian width, with $\beta$ being related to the bin size. Fitting the curves in Figure \ref{fig:modcrop} we find that as the integration radius increases both the asymptotic enhancement and the Gaussian width decrease. This implies that to optimize a small radius on the detector requires a much larger portion of the SLM to be active than in order to optimize a large target radius; which is expected given the inverse relationship between distances in the sample and detector planes.

\subsubsection{Phase Steps}
\begin{figure}
\centering
\includegraphics{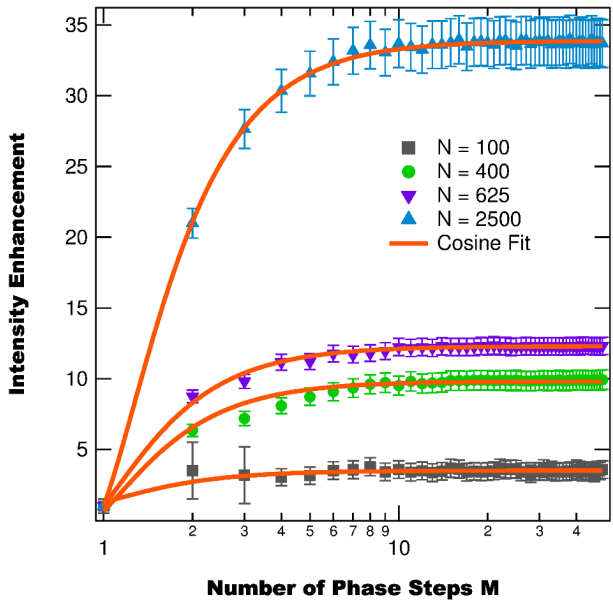}
\caption{(Color Online) Intensity enhancement as a function of the number of phase steps.}
\label{fig:dpmodel}
\end{figure}

In addition to being able to change the number of controllable channels on the SLM, we also can vary the phase resolution of each channel.  To model the effect of the number of phase steps on optimization we use an integration radius of $r=5$ $\Delta x'$, and four total bin numbers: $N=\{100,400,625,2500\}$.  Figure \ref{fig:dpmodel} shows the enhancement as a function of phase steps, which is found to quickly saturate at around 10 phase steps for each $N$ value.  Qualitatively this dependence can be understood because as the number of phase steps increases the phase resolution increases and at a certain point there will be diminishing returns in trying to attain higher phase resolutions.  

To understand this behavior we recall that the speckle pattern is an interference effect with the optimization process attempting to match the phases of different beam portions to constructively interfere \cite{Vellekoop08.01,Yilmaz13.01}.  This implies that the intensity in the target spot depends on a sum of interference terms of the form:

\begin{align}
A_n\cos(\Phi_n-\psi_n),
\end{align}
where $A_n$ is an amplitude factor, $\Phi_n$ is the phase before modulation, and $\psi_n$ is the contribution of the SLM given by
\begin{align}
\psi_n=\frac{2\pi q_n}{M},
\end{align}
with $q_n$ being an integer corresponding to the phase value giving the largest enhancement.  While the exact functionality of the intensity is a complex sum over many such terms, we find that we can fit the enhancement as a function of phase steps using only one term giving a fit function of the form:

\begin{align}
\eta=1+\eta_0\cos\left(\phi_0+\frac{\Delta\phi}{M}\right),
\label{eqn:modeldp}
\end{align}
where $1+\eta_0$ is the asymptotic enhancement and  $\phi_0$, $\Delta \phi$ are parameters which determine the shape of the function.

\subsubsection{Target Radius}
\begin{figure}
\centering
\includegraphics{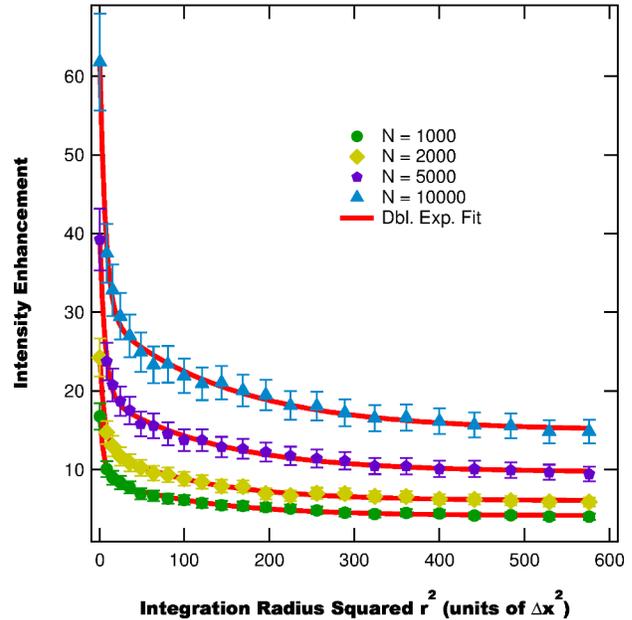}
\caption{(Color Online) Modeled intensity enhancement as a function of squared integration radius for four different bin numbers.  The enhancement is found to follow a double exponential decay.}
\label{fig:modradii}
\end{figure}

Thus far we have only considered how changing the SLM properties affects optimization.  However, we also can control the detector's parameters; most importantly, we can change the target integration radius.  In order to model the effect of the target radius on enhancement we use $M=10$ phase steps and four different bin sizes such that $N=\{1000,2000,5000,10000\}$.  Figure \ref{fig:modradii} shows the modeled enhancement which follows a double exponential as a function of the squared integration radius:

\begin{equation}
\eta =1+ A_1e^{-r^2/\sigma_1^2}+A_2e^{-r^2/\sigma_2^2}\label{eqn:r}
\end{equation}
where $A_1$,$A_2$ are amplitude factors, and  $\sigma_1$, $\sigma_2$ are Gaussian widths. 

While the decrease in enhancement with increasing target area is expected, the functional form is surprising.  To demonstrate this, we derive the expected functional form by recalling that there is a finite amount of power, $P_0$, that can be focused into the integration area. Assuming perfect enhancement -- in which all the power is focused into the target -- we would expect the maximum enhancement for a given radius to be
\begin{equation}
\eta_{max}(r)=\frac{1}{\langle I_0 \rangle}\frac{P_0}{\pi r^2},
\label{eqn:rint}
\end{equation}
where $\langle I_0 \rangle$ is the average intensity before enhancement.  However, the RPGBM results are found to follow Equation \ref{eqn:r} and not Equation \ref{eqn:rint}.  Currently the underlying physical principle determining this behavior is unknown and is an area of active research.

\subsubsection{Beam Diameter}
\begin{figure}
\centering
\includegraphics{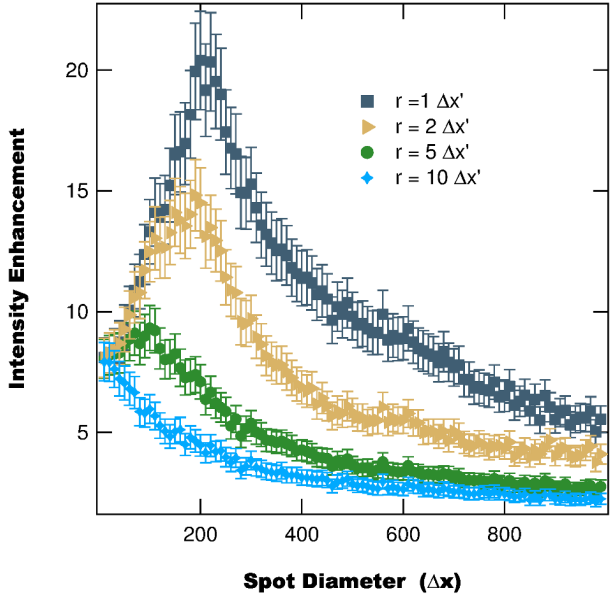}
\caption{(Color Online) Modeled intensity enhancement as a function of beam diameter.  The enhancement is found to be a peaked function, with the peak location being inversely related to the integration radius used.  This result is consistent with the Fourier relationship between the sample and detector planes.}
\label{fig:modeldiam}
\end{figure}

The last system parameter we consider is the beam diameter. To model the effect of the beam diameter on the enhancement we use $M=10$ phase steps, four different integration radii, and different bin sizes such that the beam diameter is always 10 bins (i.e. $b=1$ for a beam diameter of 10, $b=2$ for a diameter of 20, etc.). Figure \ref{fig:modeldiam} shows the enhancement as a function of beam diameter for different integration radii.  The enhancement is found to follow a peaked function where the peak location is dependent on the integration radius used.   As the integration radius decreases the beam diameter corresponding to peak enhancement is found to increase.  This suggests an inverse relationship between the beam diameter and target spot size, which is consistent with the Fourier relationship between the two planes.

\subsection{Model Summary}
We model the process of SLM based transmission optimization using a beam propagation model based on a Gaussian beam with a random phase profile.  Using the model and a sequential bin-by-bin optimization algorithim  we optimize the diffracted pattern in a target area for varying systematic parameters.  We find that the optimization depends on all parameters tested which include: bin size, active SLM area, total number of phase steps, detector integration radius, and on-sample spot size.  These results are different than those of previous models, which predict the enhancement to only depend on the number of modulated SLM channels (bins). \cite{Vellekoop07.01,Yilmaz13.01}.

In addition to our model predicting that the enhancement depends on more parameters, it also predicts a different dependence on the number bins used. To derive our systems dependnece on the number of bins we compare Equations \ref{eqn:modelbin} and \ref{eqn:Lmodel}. From these equations we find that the scale factors $\alpha$ and $\beta$ relate the two equations with $\alpha=L/\Delta L$ and $\beta=b_0/b$.  Substututing the definitions of $\alpha$ and $\beta$ into Equations \ref{eqn:modelbin} and \ref{eqn:Lmodel} we find that the model predicts an intensity enhancement dependence on SLM bin size and active SLM area as:
\begin{align}
\eta=1+\eta_0\exp\left\{-\left(\frac{Lb_0}{b\Delta L}\right)^2\right\},
\label{eqn:mbsum}
\end{align}
where $b_0$ is found to be independent of the number of phase steps and $\Delta L$ is found to decrease with increasing integration radius. Recalling that the number of bins is given by $N=(L/b)^2$, we can rewrite Equation \ref{eqn:mbsum} in terms of the number of bins,
\begin{align}
\eta=1+\eta_0\exp\left\{-\frac{N}{N_0}\right\},
\label{eqn:mnexp}
\end{align}
where $N_0=(\Delta L/b_0)^2$.  The bin number dependence in Equation \ref{eqn:mnexp} is drastically different than predicted by previous models \cite{Vellekoop07.01, Yilmaz13.01}.  We hypothesize that this difference arises due to the inclusion of beam propagation effects and enhancement saturation due to detector noise.

The other difference between the RPGBM and previous models, is that the RPGBM predicts that the number of SLM phase steps, detector integration radius, and the on-sample beam spot size also affect the intensity enhancement.  The dependence on the number of phase steps arises due to optimization being related to controlled interference, while the influence of the integration radius and on-sample beam spot size occurs due to the diffractive nature of the enhancement phenomenon.  Since the sample and detector planes are related via a Fourier Transform the effect of changing distances in one plane directly affects distances in the other plane.

\textcolor{black}{Finally, one of the major simplifications of RPGBM is using a smooth Gaussian amplitude for the scattered beam, whereas real scattering results in both the amplitude and phase of the beam being modulated. Experimentally we find that the amplitude of the beam leaving the scattering sample is approximated by a Gaussian with an additive noise term. When performing calculations we find that adding a noise term has negligible effect on the functional form of the optimization.  This is due to the model having the SLM only affecting the phase of the scattered beam and not its amplitude.  In reality, however, the scattered beam's phase and amplitude are coupled due to the material and we expect that using an SLM to change the incident beam's phase will result in a small change in the scattered amplitude.   To account for this effect we are currently extending the RPGBM with a model of scattering based on transmission eigenchannels and random-matrix theory \cite{Beenakker97.01,Vellekoop08.01,
Vellekoop08.02,Mello88.01}.}

\section{Experimental Method}
\begin{figure}
\centering
\includegraphics{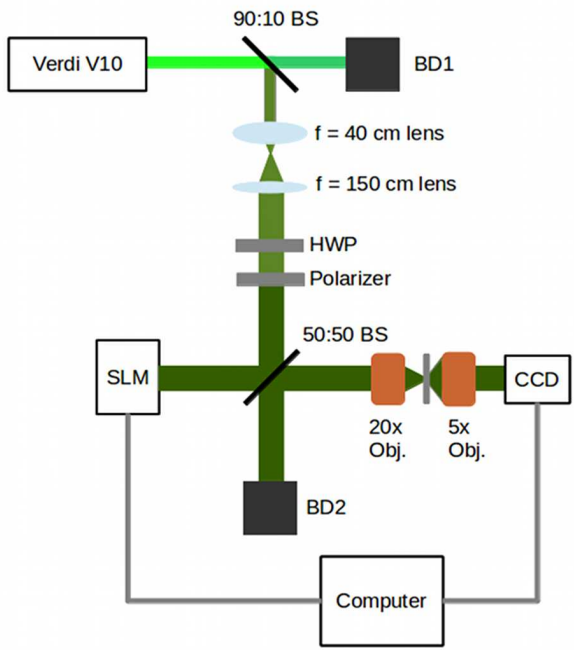}
\caption{(Color Online) Optical setup schematic.} 
\label{fig:schematic}
\end{figure}

We experimentally measure the effect of various system parameters on optimization using a controlled transmission optical setup, which consists of a high-speed LCOS SLM from Boulder Nonlinear Scientific, a Coherent Verdi V10 Nd:YVO$_4$ laser, a high speed Thorlabs CMOS camera \textcolor{black}{(8-Bit,pixel size of 5.2 $\mu$m, variable exposure time from 37 $\mu$s to 2 ms)}, and various focusing and polarization optics. Figure \ref{fig:schematic} shows a schematic of the system.

The beam from the laser is sampled by a 90:10 beamsplitter (BS) and then expanded by a factor of 3.75$\times$ and passed through a half-waveplate (HWP) polarizer pair to control beam intensity.  The expanded beam is then reflected by a beamsplitter onto the SLM which modulates the phase of the laser beam.  After modulation the beam is focused onto the sample using a 20$\times$ high working distance (HWD) objective ($WD=20$ mm), with the scattered light being collected onto a CMOS camera by a 5$\times$ HWD objective ($WD=37.5$ mm).

To keep optimization times managable we bin the SLM pixels into $N$ bins with each bin having an edge size of $b=L/N$, where $L$ is the total number of active pixels on a side.  The bins are optimized using a sequential optimization algorithm \cite{Vellekoop07.01,Vellekoop08.01} in which each bin's phase value is updated through $M$ phase steps of size $\Delta \phi=2\pi/M$.  After each update the camera takes an image which is used to calculate the intensity within the target area.  After all $M$ steps are completed, the bin's phase is fixed to the phase value corresponding to the largest intensity measured.  This procedure repeats for all bins until an optimized phase pattern is displayed on the SLM.  Using this algorithm with our setup we achieve \textcolor{black}{iterative} rates of 160-180 Hz.

\section{Results and Discussion}
To characterize the controlled transmission setup we systematically vary the five system parameters: bin size, active SLM area, number of phase steps, target area, and beam spot size. We measure each dependence by varying one parameter, while holding all other parameters fixed, and measure the intensity enhancement.
To obtain better statistics we perform ten optimization runs for each parameter set and find the average enhancement and error from the ten runs.

In order to separate which effects are due to the systematic parameters and which are due to the opaque sample, we perform measurements on five different sample types:  ZrO$_2$ NP-doped polyurethane (PU), ZrO$_2$ NP-doped polyepoxy(PE), Y$_2$O$_3$ pressed ceramic, ground glass, and printer paper.  From these measurements we find that the functional form of the enhancement as a function of system parameters is independent of sample type, with the different samples only affecting fit parameters (e.g. peak enhancement, shape parameters).  This suggests that the measured dependencies are a function of the optical setup and not the samples.  Since the functional forms are consistent across samples; we present in the following sections the enhancement measured using ground glass, as it's speckle pattern is found to be the most stable over time and it produces the largest enhancements.

\subsection{Bin Size}
\begin{figure}
 \centering
 \includegraphics{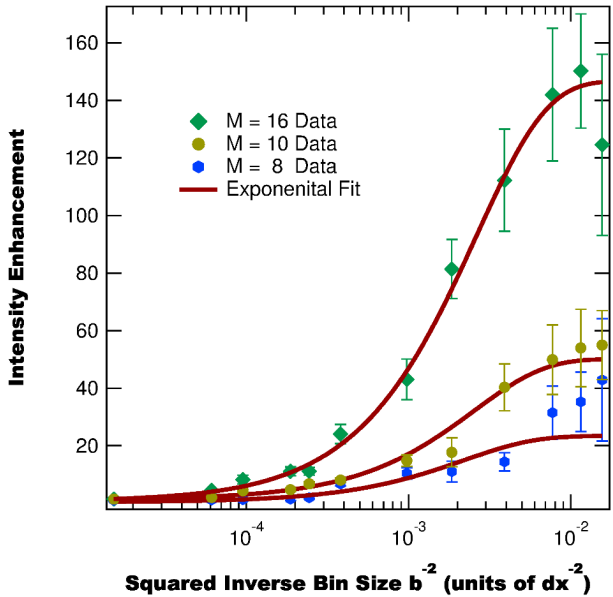}
 \caption{(Color Online) Intensity enhancement as a function of squared inverse bin size. The enhancement is found to follow an exponential function, which is consistent with the RPGBM results.}
 \label{fig:expetab}
\end{figure}

The first parameter we investigate is the bin size.  We measure the intensity enhancement at 12 different bin sizes using a beam spot of diameter $\approx$ 350 $\mu$m, an integration radius of 2 px, and three different total number of phase steps, $M=\{8,10,16\}$.  Figure \ref{fig:expetab} shows the enhancement as a function of squared inverse bin size, which is found to follow a function of the form

\begin{equation}
 \eta=1+\eta_0(1-e^{-b_0^2/b^2}),\label{eqn:bins}
\end{equation}
where $1+\eta_0$ is the asymptotic enhancement and $b_0$ determines the enhancement's shape. We find the parameters $\eta_0$ and $b_0$ for each $M$ value by fitting the curves in Figure \ref{fig:expetab}, which are tabulated in Table \ref{tab:bin}.  The asymptotic enhancement is found to increase with the number of phase steps, while the shape factor, $b_0$, is found to be constant within uncertainty.  These results are functionally consistent with the RPGBM results.  The difference in the magnitude of enhancement between model and experiment is due to three factors: 1) imperfect matching of parameters between experiment and modeling, 2) different noise levels, and  3) a divergence between the model and experiment related to the number of phase steps (which is discussed later in the paper).

\begin{table}
\centering
\begin{tabular}{|c|c|c|}
\hline
$M$  &  $\mathbf{\eta_0}$  & $b_0$  \\ \hline
8       &   $23.4 \pm 4.9$  &  $21.4 \pm 3.5$ \\
10    &   $50.1 \pm 7.2$  &   $19.9  \pm 1.9$\\
16  &   $146 \pm 14$   &  $19.5 \pm 1.3$\\ \hline
\end{tabular}
\caption{Fit parameters from Equation \ref{eqn:bins} for the intensity enhancement as a function of bin size.}
\label{tab:bin}
\end{table}

\subsection{SLM Cropping}
\begin{figure}
\centering
\includegraphics{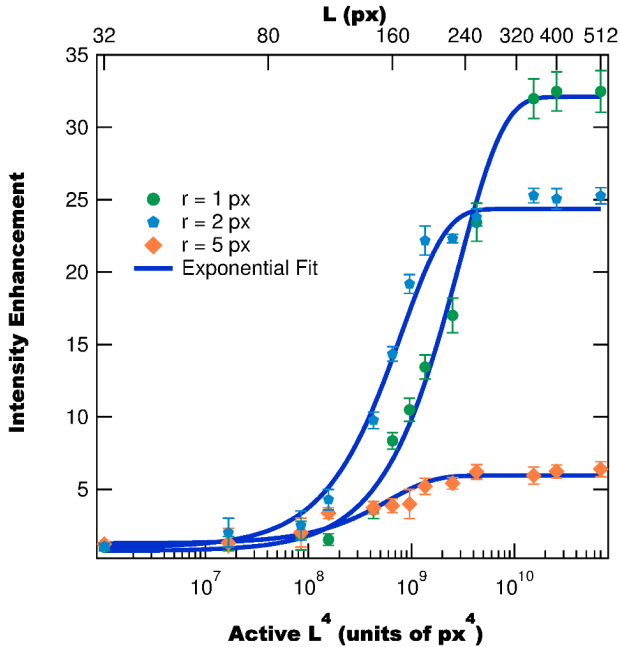}
\caption{(Color Online) Measured intensity enhancement as a function of the quartic active side length, $L^4$.  The enhancement is found to follow an exponential function, which is different than predicted by the RPGBM. }
\label{fig:crop}
\end{figure}

Next we test the effect of changing the active SLM area using a spot diameter of $\approx$ 200 $\mu$m, a bin size of $b=8$ px, $M=8$ phase steps, and three different integration radii, $r=\{1,2,5\}$. We first measure the enhancement with the full SLM active, after which we ``shut off'' the outer rows/columns -- such that the active area is always a centered square -- and perform optimization again.  We continue to do this until only a 4 bin $\times$ 4 bin (32 px $\times$ 32 px) area remains active.  Figure \ref{fig:crop} shows the intensity enhancement as a function of quartic active side length, $L^4$.  From the figure we find that the enhancement, as a function of active length, $L$, is found to behave functionally as

\begin{equation}
\eta=1+\eta_0\left[1-\exp\left\{\left(\frac{L}{\Delta L}\right)^4\right\}\right],\label{eqn:L}
\end{equation}
where $\Delta L$ is a width parameter and $1+\eta_0$ is the asymptotic enhancement.  From Equation \ref{eqn:L} we find that the explicit dependence of the enhancement on active side length is different than predicted by RPGBM (i.e. Gaussian in $L$ for the RPGBM and Gaussian in $L^2$ for experiment).  

A possible explanation for this discrepancy is related to how the RPGBM treats SLM cropping versus the real world implementation.  In the RPGBM, cropping of the active SLM area is implemented by shutting off modulation in grid points that represent the sample's {\em exit} plane. However, in reality SLM cropping shuts off modulation of portions of the light {\em incident} on the sample. The modulated light incident on the sample is then transmitted through the sample with various spatial components interfering.  This interference is not accounted for in the RPGBM, which could lead to the divergence between the RPGBM and experiment.

Despite the explicit dependence on $L$ being different, the general dependence -- enhancement increasing to a constant value as $L$ increases-- is consistent. Additionally, we find that the behavior of the enhancement's dependence as the integration radius changes is also consistent.  To demonstrate this consistency we fit Figure \ref{fig:crop} to Equation \ref{eqn:L} and find $\eta_0$ and $\Delta L$ for each integration radius tested.  Table \ref{tab:crop} compiles the fitting results.  Both the amplitude, $\eta_0$, and width parameter, $\Delta L$, are found to decrease with increasing integration radius, which is predicted by the RPGBM.

\begin{table}
\centering
\begin{tabular}{|c|c|c|}
\hline
$r$  &  $\mathbf{\eta_0}$  & $\Delta L$  \\ \hline
1       &   $32.10 \pm 0.80$  &   $232.6 \pm 3.6$ \\
2    &   $24.36 \pm 0.25$ &   $168.3  \pm 1.9$\\
5  &   $5.94 \pm 0.11$   &  $156.4 \pm 3.6$\\ \hline
\end{tabular}
\caption{Fit parameters from Equation \ref{eqn:L} for the intensity enhancement as a function of active area. Both $\eta_0$ and $\Delta L$ decrease with increasing integration radius, which is consistent with the RPGBM.}
\label{tab:crop}
\end{table}

\subsection{Phase Steps}
The final SLM parameter we vary is the number of phase steps used during optimization. For these measurements we use a spot diameter of $\approx$ 350 $\mu$m, an integration radius of 2 px, and three bin sizes: $b=16$ px, $b=32$  px, and $b=64$ px. We find that the enhancement depends on the number of phase steps, $M$, as:

\begin{align}
\eta=1+\eta_0\cos^p\left(\frac{\pi}{2M}\right),
\label{eqn:Mdep}
\end{align}
where $1+\eta_0$ is the asymptotic enhancement and $p$ is an exponent which controls the shape of the function. While the RPGBM predicts that $p=1$, from fitting we find that $p>1$ and decreases as $N$ increases, as shown in Table \ref{tab:expM}.  This result is unexpected and the underlying mechanism is currently unknown.

One possible explanation is that realistic samples may complicate the coupling between the modulated phase incident on the sample and the phase exiting the sample such that only a fraction of the light exiting the sample has an optimized phase.  The result of having less control would be to decrease the enhancement, which is consistent with $p>1$.  Also, we would expect effects due to the sample to decrease as $N$ increases, since the size of the modulated area decreases and gives a greater control over transmission through the sample.  This increased control would cause $p$ to decrease, which we see experimentally.

\begin{figure}
 \centering
 \includegraphics{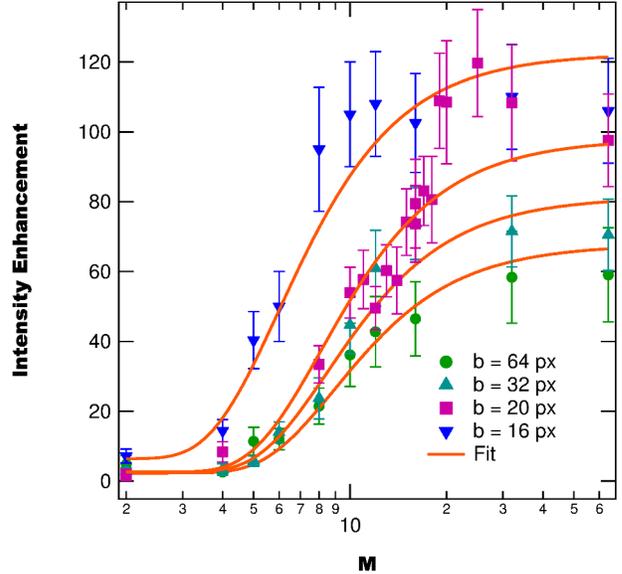}
 \caption{(Color Online) Intensity enhancement as a function of phase steps.}
 \label{fig:expetap}
\end{figure}

\begin{table}
\centering
\begin{tabular}{|c|c|c|}
\hline
$N$  &  $\mathbf{\eta_0}$  & $p$  \\ \hline
64       &   $65.5 \pm 6.4$  &  $61.0  \pm  5.3$ \\
256    &   $78.8 \pm 6.4$  &   $59.2 \pm 6.5$\\
625    &    $95.3 \pm 4.1$          & $53.5\pm8.6$\\
1024  &   $116.0\pm 8.2$   &  $29.0 \pm 3.5$\\ \hline
\end{tabular}
\caption{Fit parameters from Equation \ref{eqn:Mdep} for the intensity enhancement as a function of the number of phase steps.}
\label{tab:expM}
\end{table}

\subsection{Target Radius/Area}
\begin{figure}
 \centering
 \includegraphics{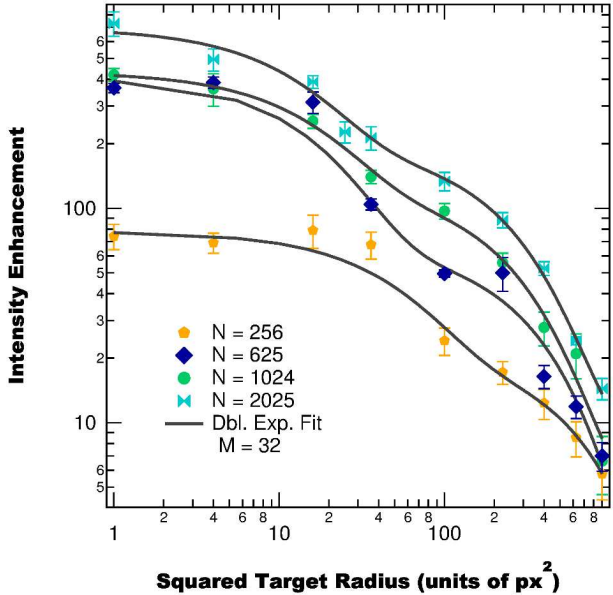}
 \caption{(Color Online) Intensity enhancement as a function of integration radius measured using ground glass with 32 phase steps, a spot size of 380 $\mu$m and four different total number of bins.  The enhancement is found to behave as a double exponential which is consistent with the RPGBM.}
 \label{fig:exprad}
\end{figure}

Thus far we have only considered the effects of SLM parameters.  At this point we turn our attention to the enhancement's dependence on target radius. We measure the intensity enhancement using a beam spot size of 380 $\mu$m, $M=32$ phase steps, four different numbers of bins, $N=\{256,625,1024,2025\}$, and nine integration radii/areas.  The measured enhancement as a function of integration radius, shown in Figure \ref{fig:exprad},  is found to behave as the sum of two Gaussians. This behavior is identical to the RPGBM.

We also determine amplitudes and Gaussian widths as a function of $N$ by fitting the measured enhancement as a function of integration radius to Equation  \ref{eqn:r}. Table \ref{tab:area} compiles the fit results.  The amplitudes are found to increase with bin number -- consistent with our other results -- and the widths are found to decrease as the bin number increases.  Additionally, the Gaussian widths appear to reach a constant value as the number of bins increases, with the widths for $N=1024$ being within uncertainty of those for $N=2025$.  

As with the RPGBM model's results, the underlying mechanism behind the experimental enhancement's target area dependence is currently unknown.  Given that both the model and experiment have the same functional dependence, suggests that the physical phenomenon responsible is related to beam propagation effects.  Currently we are performing further experiments and modeling in order to better understand how other parameters, besides the number of bins, affects the enhancement's dependence on integration radius.

\begin{table}
\centering
\begin{tabular}{|c|c|c|c|c|}
\hline
$N$  &  $A_1$  & $\sigma_1$  &  $A_2$  & $\sigma_2$  \\ \hline
256       &   $22 \pm 13$  &  $27.1 \pm 5.5$  & $56 \pm 12 $  &  $7.4 \pm 1.5$ \\
625    &   $67.1 \pm 3.3$  &   $19.1  \pm 1.7$  &  $344\pm 17$  &  $4.23 \pm 0.18$\\
1024  &   $129 \pm 16$   &  $15.85 \pm 1.66$  &  $312\pm 33$  &  $3.95 \pm 0.38$ \\ 
2025  &   $190 \pm 19$   &   $16.10 \pm 0.91$  &  $496 \pm 71$ &  $3.77 \pm 0.35$   \\ \hline
\end{tabular}
\caption{Fit parameters from Equation \ref{eqn:r} for the intensity enhancement as a function of integration radius.}
\label{tab:area}
\end{table}

\subsection{Spot Size}
\begin{figure}
\centering
\includegraphics{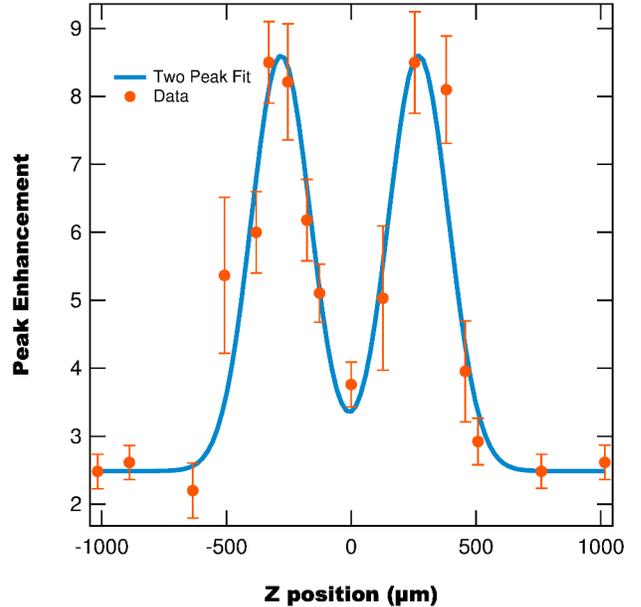}
\caption{(Color Online) Peak enhancement as a function of position along the optical axis, where $z=0$ is the focal point of the focusing lens.  The spot size at $z=0$ is $\approx 0.9$ $\mu$m.}
\label{fig:zpos}
\end{figure}

The last system parameter we vary is the on-sample beam spot size.  To measure the enhancement's spot size dependence we use $N=1024$ bins, $M=16$ phase steps, and integration radii of 2 px, 20 px, 30 px. We first measure the enhancement with the sample positioned within the focal length of the focusing objective such that the on-sample beam diameter is 600 $\mu$m.  After the initial measurement we systematically translate the sample and measure the enhancement at fixed $z$ positions until the sample translates through the focal point and reaches an on-sample beam diameter of 600 $\mu$m again.   Figure \ref{fig:zpos} shows the peak enhancement as a function of position along the optical axis, where $z=0$ is the focal point of the focusing lens. From Figure \ref{fig:zpos} we find that the enhancement is symmetric about the focal point, with the peak value occurring at a nonzero distance from the focal point.  

\begin{figure}
 \centering
 \includegraphics{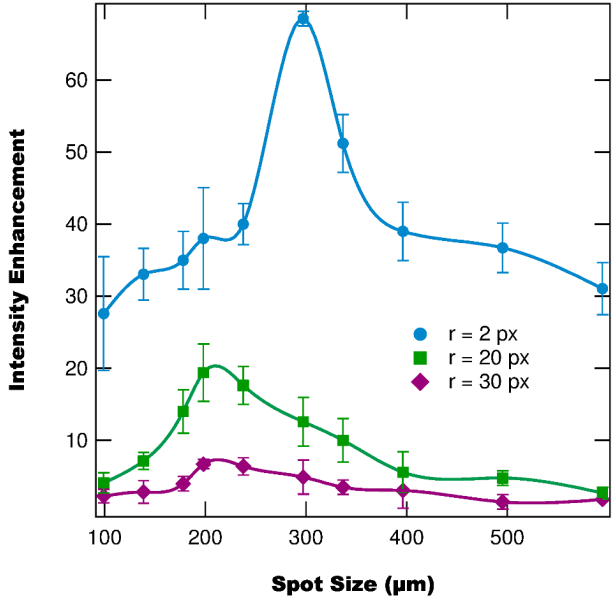}
 \caption{(Color Online) Measured intensity enhancement as a function of the on-sample spot size.  The enhancement is found to follow a peaked function with the peak location being inversely dependent on the integration radius, consistent with the RPGBM result.}
 \label{fig:etadiam}
\end{figure}

We can convert the $z$-position into the spot diameter, $2w$,  using ray matrix Gaussian beam propagation.   Assuming that the beam incident on the focusing lens is at/near it's waist, the Gaussian width, $w$, at position $z$ is given by:

\begin{equation}
w(z)=\frac{1}{kfw_0}\sqrt{4f^2(f+z)^2+k^2w_0^4z^2}
\label{eqn:width}
\end{equation}
where $w_0$ is the beam diameter at the focusing lens, $f$ is the focal length of the lens, and $k=2\pi/\lambda$ where $\lambda$ is the wavelength of light.  Note that Equation \ref{eqn:width} is symmetric about the focusing lens's focal point. Therefore we average the enhancement measured on both sides of the focal point to find the intensity enhancement as a function of spot diameter which is shown in Figure \ref{fig:etadiam}.  

The experimentally measured intensity enhancement is found to peak at a nonzero spot diameter, with the diameter corresponding to peak enhancement increasing as the integration radius decreases.  Additionally the width of the peaked function is found to decrease as the integration radius decreases.  These behaviors are consistent with the prediction of the RPGBM.

\subsection{Different Samples}
The experimental enhancement's dependence on the different systematic parameters diverges from both previous models (Equation \ref{eqn:Vmodel} and \ref{eqn:Ymodel}) \cite{Vellekoop07.01,Yilmaz13.01} and the RPGBM.  While the RPGBM predicts similar behavior to experiment, some of the dependencies are functionally different.  The most likely source of divergence between experiment and the RPGBM is the model's treatment of the sample as a ``black box''. This treatment predicts that the enhancement is independent of the sample properties (e.g. sample thickness and scattering length).  

While a precise characterization of the enhancement's dependence on sample parameters is beyond the scope of this paper, we consider a simple direct comparison between six different samples: paper, ground glass, Y$_2$O$_3$ ceramic, and three formulations of ZrO$_2$ NP-doped polymers.  We perform optimization using the same experimental parameters ($b=16$ px, $M=32$, $r=2$ px, $w=250$ $\mu$m) at five different points on each sample to find the spatially averaged intensity enhancement, which is tabulated in Table \ref{tab:samp}. From Table \ref{tab:samp} we see a wide variation in enhancement when using different samples, with the largest enhancement being 36 $\times$ larger than the smallest.  

\begin{table*}
\centering
\begin{tabular}{|c|c|c|c|}
\hline
Sample  & Thickness ($\mu$m)  & Scattering Length ($\mu$m)  &  Enhancement \\ \hline
Ground Glass  & $1564  \pm  75$ & $970.7 \pm 2.1$  & $172 \pm 12$ \\
Paper  & $85.1\pm6.4$ & $(2.654 \pm 0.026)\times 10^{-3}$  &$9.4 \pm 1.0$ \\
10 wt\% ZrO$_2$ NP/PU & $867 \pm 67$ & $4.11  \pm  0.28$   &$5.50\pm0.45$ \\
10 wt\% ZrO$_2$ NP/PE &   $1036 \pm 50$& $3.9 \pm  1.2$   &$4.79\pm0.49$ \\
1 wt\% ZrO$_2$ NP/PU &     $959\pm37$& $50.8  \pm  3.1 $     &$44.5 \pm 1.2$\\
Y$_2$O$_3$ Ceramic   &  $358 \pm 44$ & $(2.944 \pm 0.029)\times 10^{-2}$  &  $12.8  \pm  1.3$  \\ \hline
\end{tabular}
\caption{Maximum enhancement obtained for different samples using system parameters of $b=16$ px, $M=32$, $r=2$ px, $w=250$ $\mu$m.  There is a 36$\times$ difference between the smallest and largest enhancement.}
\label{tab:samp} 
\end{table*}

We deduce several possible factors from these preliminary measurements which may affect the enhancement: the sample persistence time and scattering length.  The first factor, the persistence time, is a measure of how long a sample will produce the same speckle pattern; which directly affects how well an SLM system can optimize transmission \cite{Vellekoop07.01,Vellekoop08.01}.  From our measurements we find that the ground glass and NP samples have stable speckle patterns over a period of days, while the paper's speckle pattern changes in tens of minutes.  This results in paper  having a relatively low enhancement despite being the thinest sample.  While the persistence time is important to optimization, it does not explain the wide variation in the most stable samples.  For these samples we see that the enhancement is largest for large scattering lengths (ground glass) and smaller for small scattering lengths (NP samples).  

The result linking the enhancement to the scattering length is important to the long-term goals of our study, which are to use optimal transmission as a method of authenticating NP-doped polymeric PUFs.  With this in mind we note that for the two different 10 wt\% ZrO$_2$ NP-doped polymers the enhancements are within uncertainty of each other, despite the samples having different host polymers and different thickness. Additionally, comparing 1 wt\% and 10 wt\% ZrO$_2$ NP-doped PU, we find that the 1 wt\% enhancement is almost 10 $\times$ as large as for the 10 wt\%.  These results suggest that the polymeric hosts have a negligible effect on the optimization process, while the particle concentration has a large inverse effect.  This can be understood according to Mie's scattering theory which predicts that the scattering length -- for a system of scattering spheres -- is inversely proportional to the scatterer density \cite{Mie08.01}. This implies that as the concentration increases the number of scattering 
events increases. We hypothesize that this increase in scattering events decreases our ability to control transmission of light through the system.  While the exact mechanism for this effect is unknown, Mosk has proposed that by increasing the number of scattering events, either by increasing thickness or concentration, will result in noise playing a larger effect in optimization \cite{Mosk14.01}, resulting in the enhancement decreasing \cite{Yilmaz13.01}.  Further work is required to better understand this effect.

%\subsection{Phenomenological Model}
%\textcolor{From testing optimization as a function of different system variables we have determined several equations to describe the functionality of the enhancement.  Ideally we would be able to combine }

\section{Conclusion}
We systematically measure the dependence of optimal transmission on five different system variables: SLM bin size, number of SLM phase steps,  active SLM area, detector integration radius, and the on-sample beam spot size.  From our measurements we find optimization to depend on all five system variables as well as the characteristics of the sample used for optimization.  These results are contradictory to previous models of universal optimal transmission which proposed that optimization is only dependent on the number of SLM channels used \cite{Vellekoop07.01,Vellekoop08.02} or the number of SLM channels and the signal-to-noise ratio \cite{Yilmaz13.01}.  To understand the nature of these contradictory results we develop a model based on the propagation of a Gaussian beam with a random phase profile.  

We find that the model is mostly consistent with experimental results, with the effects of beam propagation primarily arising due to the Fourier relationship between the sample and detector planes. This relationship leads to the beam diameter and active SLM area being inversely related to the target area to be optimized. This implies that to best optimize a small target radius a large beam spot size and large active SLM area are required, while to optimize a large target area requires less of the SLM to be active and a smaller spot size.  We also find that including noise effects into the model leads to the correct enhancement dependence on the number of bins.

While the RPGBM and experiment are mostly consistent, there are still some key variations: namely, the enhancement's dependence on the number of phase steps and active SLM area. These deviations most likely arise due to the RPGBM's treatment of the sample \textcolor{black}{and scattering}, which assumes that the sample does not affect the optimization process \textcolor{black}{and that the scattered phase and amplitude are independent}. These assumptions, however, are incorrect as we see from experiment that the sample does affect the optimization process (primarily due to the persistence time and the concentration of scatterers) \textcolor{black}{and that the scattered amplitude and phase are correlated by scattering within the sample}.  To address these shortcomings we currently are in the process of performing systematic studies of how different sample properties affect the enhancement \textcolor{black}{as well as working on extending the RPGBM to more accuaretly represent scattering within the sample}. 
\textcolor{black}{To extend the RPGBM we are utilizing two different computational methods to describe scattering: random matrix transmission eigenchannels \cite{Beenakker97.01,Vellekoop08.01,Vellekoop08.02} and monte carlo scattering simulations \cite{Wang95.01,Wang97.01,Alerstam08.01,Fang09.01}. The scattering models will be implemented into the RPGBM as follows: for each phase mask tested the beam will be passed through the scattering algorithim to determine the beam amplitude and phase exiting the sample. As both algorithims depend on many iterations using random numbers we will run the same phase mask multiple times to determine an ensemble average amplitude and phase.  These ensemble averaged exit beams will then be transformed using Equation \ref{eqn:modfield} to determine the intensity in the detector plane.  This process will be repeated according to the optimization algorithim until the optimal phase mask is determined.}

\acknowledgments
%\section{Acknowledgements}
This work was supported by the Defense Threat Reduction Agency, Award \# HDTRA1-13-1-0050 to Washington State University.

\bibliographystyle{osajnl}
\bibliography{ASLbib}

\end{document}